# Disentangling stress and curvature effects in layered 2D ferroelectric CuInP$_2$S$_6$

Yongtao Liu,[1*] Anna N. Morozovska,[2] Ayana Ghosh,[3] Kyle P. Kelley,[1] Eugene A. Eliseev,[4] Jinyuan Yao,[5] Ying Liu,[5] and Sergei V. Kalinin[6*]

[1] Center for Nanophase Materials Sciences, Oak Ridge National Laboratory, Oak Ridge, TN 37922, United States

[2] Institute of Physics, National Academy of Sciences of Ukraine, 46, pr. Nauky, 03028 Kyiv, Ukraine

[3] Computational Sciences and Engineering Division, Oak Ridge National Laboratory, Oak Ridge, TN 37830, USA

[4] Institute for Problems of Materials Science, National Academy of Sciences of Ukraine, 3, Krjijanovskogo, 03142 Kyiv, Ukraine

[5] Department of Physics, Pennsylvania State University, University Park, PA 16802, USA

[6] Department of Materials Science and Engineering, The University of Tennessee, Knoxville, TN 37996, United States

Nanoscale ferroelectric 2D materials offer unique opportunity to investigate curvature and strain effects on materials functionalities. Among these, CuInP$_2$S$_6$ (CIPS) has attracted tremendous research interest in recent years due to combination of room temperature ferroelectricity, scalability to a few layers thickness, and unique ferrielectric properties due to coexistence of 2 polar sublattices. Here, we explore the local curvature and strain effect on the polarization in CIPS via piezoresponse force microscopy and spectroscopy. To explain the observed behaviors and decouple the curvature and strain effects in 2D CIPS, we introduce finite element Landau-Ginzburg-Devonshire model. The results show that bending induces ferrielectric domains in CIPS, and the polarization-voltage hysteresis loops differ in bending and non-bending regions. Our simulation indicates that the flexoelectric effect can affect local polarization hysteresis. These studies open a novel pathway for the fabrication of curvature-engineered nanoelectronic devices.

*Corresponding emails: liuy3@ornl.gov; sergei2@utk.edu

Within two decades since the work by Geim and Novoselov in 2004,[1] 2D materials including graphene and transition metal dichalcogenides[2] have become one of the central topics of research in condensed matter physics, quantum materials, and electronics. In addition to the intrinsic excellent physical properties of 2D materials, e.g., high electron mobility[3] and thermal conductivity[4], new phenomena have been discovered in the twisted bilayers as driven by emergent electronic instabilities.[5-8] Over the last several years interest shifted towards the layered materials that support additional functionalities including ferromagnetic[9] and ferroelectric[10], this opening the route to ferroelectricity at atomic thicknesses. Since 2D materials can withstand significantly large curvature and strain, these materials systems allow to get insight into novel phenomena based on the curvature[11] and strain effects.[12,13] With the coupling between polarization and strain effect being significant even for 3D ferroelectrics, this is expected to be even more pronounced in 2D ferroelectric materials that can support larger curvatures.[14-18]

Of these systems, $CuInP_2S_6$ (CIPS) was discovered to exhibit ferroelectricity at room temperature,[10,19-21] making it highly promising for advanced applications utilizing ferrielectricity and antiferrielectricity down to the limit of a single layer.[19,22-24] Ferroelectricity, the equivalent of ferrimagnetism, can be termed as an antiferroelectric order, but with a switchable spontaneous polarization created by two sublattices with spontaneous dipole moments that are antiparallel and different in magnitude.[25]

In the context of the 2D material, of particular interest is the control of polarization via stress and curvature. The curvature is expected to directly contribute to flexoelectric field in the material and can also affect the ferroelectric phase stability. At the same time, stress is one of the well-recognized control parameters in ferroelectric thin films that can couple to polarization and induce transitions between structural variants. For Cu-based layered chalcogenides, stress-induced phase transitions[26] and strain engineering[14,16] become especially important in ultrathin films of $CuInP_2(S,Se)_6$,[10,27,28] Despite the significant fundamental and practical interest in bulk[29] and nanosized $CuInP_2(S,Se)_6$,[30] the influence of stress and strains on the local switching of its spontaneous polarization is explored only weakly. From theoretical perspective, there has been very little effort on both mesoscopic and atomistic levels of theory. At the same time, from an experimental perspective, while curvature of in 2D materials can be measured directly from observed topography, the strain state is determined by the non-observable interactions such as adhesion to the substrate, friction at the film-substrate interface, and sample history.

Previously, mechanical modulation of polarization in CIPS thin film was discovered. Chen et al. was able to artificially generate large-scale stripe domains of hundreds of microns by corrugating the CIPS flake, where it is believed that the giant strain gradient is introduced via high curvature.[16] Similar curvature induced domains were observed by several other groups.[14,18] Similarly, mechanical switching of polarization in CIPS has been demonstrated.[14,18] This effect was attributed to flexoelectricity owing to the formation of giant strain gradient. Almost at the same time, another work revealed strong second harmonic generation (SHG), ~160-fold enhancement compared to unstrained region, of wrinkle nanostructures in CIPS, this SHG enhancement can be modulated by the applied strain.[17] Angle resolved SHG also revealed a distorted pattern which was attributed to a photoelastic effect.[17]

Here, we decouple the curvature and strain effects in the 2D CIPS nanoflakes using the combination of scanning probe microscopy with mesoscopic and finite element method, and LGD simulation. We show that bending induces domains in CIPS, and the polarization-voltage hysteresis loops in bending regions differ from those in non-bending regions. The simulation provides insights into the mechanism behind these observations.

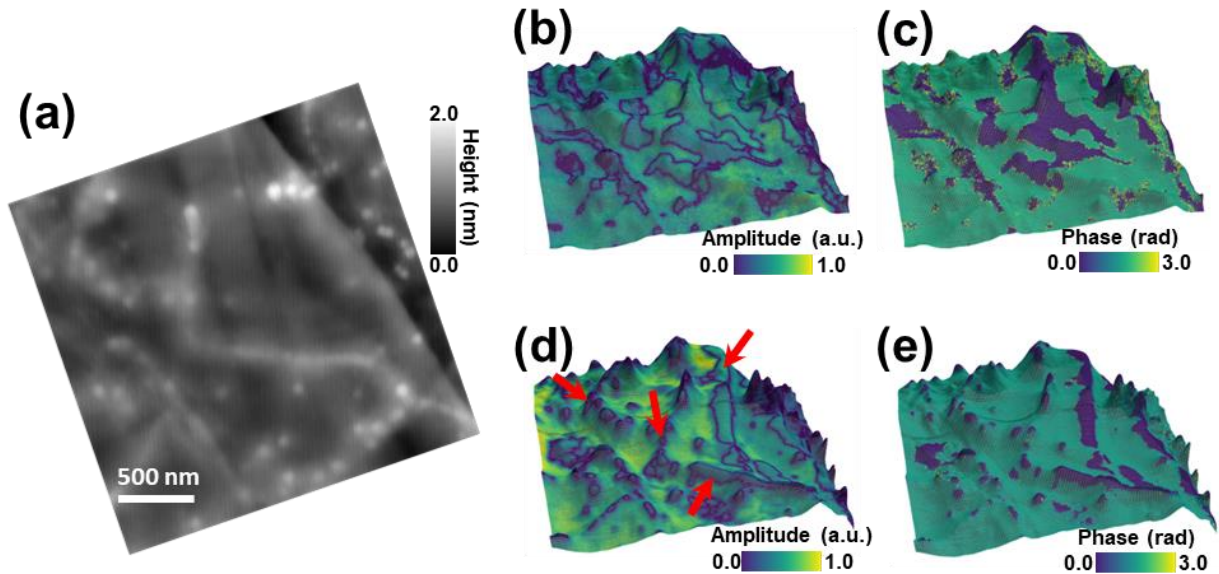

Figure 1. Band Excitation Piezoresponse Force Microscopy shows the relationship between domain evolution and morphology in CIPS. (a) Topography of a CIPS flake. (b-c) BEPFM results of original CIPS flake, (d-e) BEPFM results of the CIPS flake after stored in vacuum for 11 days. The BEPFM results in (b-e) are plotted over topography to highlight the relationship between domains and morphology, where the color represents PFM amplitude or phase and the geometry

represents topography. Obviously, after 11 days, the domain walls move towards the high curvature regions as indicated by red arrows in (d).

Single crystal $CuInP_2S_6$ was synthesized through the chemical vapor transport (CVT) method.[31] The Cu, In, P, and S powder were mixed in a stoichiometric ratio and loaded into a quartz tube (10 mm inside diameter, 18 cm length). The transport agent, 16 mg $I_2$, was added. The quartz tube was sealed under vacuum and then heated up in a horizontal double-zone furnace with the hot and cold ends set at 750 °C and 650 °C, respectively. After 14 days, the furnace was shut down, and the quartz tube was naturally cooled down to room temperature and yellow plate-like crystals were found at the cold end of the quartz tube. To bend the CIPS flakes, 30 nm Pt and 5/45 nm Ti/Au bottom electrodes were patterned on $SiO_2$ on a Si substrate by photolithography and created by electron-beam evaporation and lift-off process. CIPS flakes are mechanically exfoliated from the bulk crystal and transferred onto pre-patterned bottom electrodes by the polydimethylsiloxane (PDMS) dry transfer technique. Then, the CIPS flakes at the edge of bottom electrode are bent.

Figure 1 shows band excitation piezoresponse force microscopy (BEPFM) results of the typical ferroelectric domains of a 50 nm thick CIPS flake on Pt/Si substrate, which are very similar to previous report. [10] Shown in Figure 1a is the topography of the BEPFM measurement area, BEPFM amplitude and phase over the topography (where color represents amplitude and phase signals, and the geometry represents the topography signal) in Figure 1b and 1c, respectively, show the ferroelectric domain structure in the pristine CIPS flake. It is seen that the domain distribution is somehow correlated to the topography. However, after the CIPS flake was stored in vacuum condition for 11 days, a variation of ferroelectric domain is observed in the BEPFM results in Figure 1d-e. In comparison with the pristine domain structure in Figure 1b-c, it is obvious that the domain walls move toward the area with a larger topography variation, i.e., larger curvature area, examples are marked by red arrows in Figure 1d. This spontaneous domain wall motion induces a shrink of dark domains, as shown in Figure 1e.

The domain wall motion toward high curvature area can also be induced by electric modulation, as shown in Figure 2. Figure 2a shows the topography of the studied area. Shown in Figure 2b-c are the BEPFM amplitude and phase plotted over topography, indicating the pristine domains in this area, where color represents amplitude or phase signals and the geometry represents topography. Next, 4 V DC bias was applied through the tip and scanned over this area. Figure 2d-e shows the amplitude and phase over topography after applying 4 V DC bias. Noteworthily, for vdW ferroelectrics, electric field writing often results in leakage current and/or material breakdown, which does not guarantee effective polarization switch.[19,32,33] Nonetheless, in this work, a domain wall motion toward the high curvature area is observed, as shown in the comparison between Figure 2b-c and Figure 2d-e.

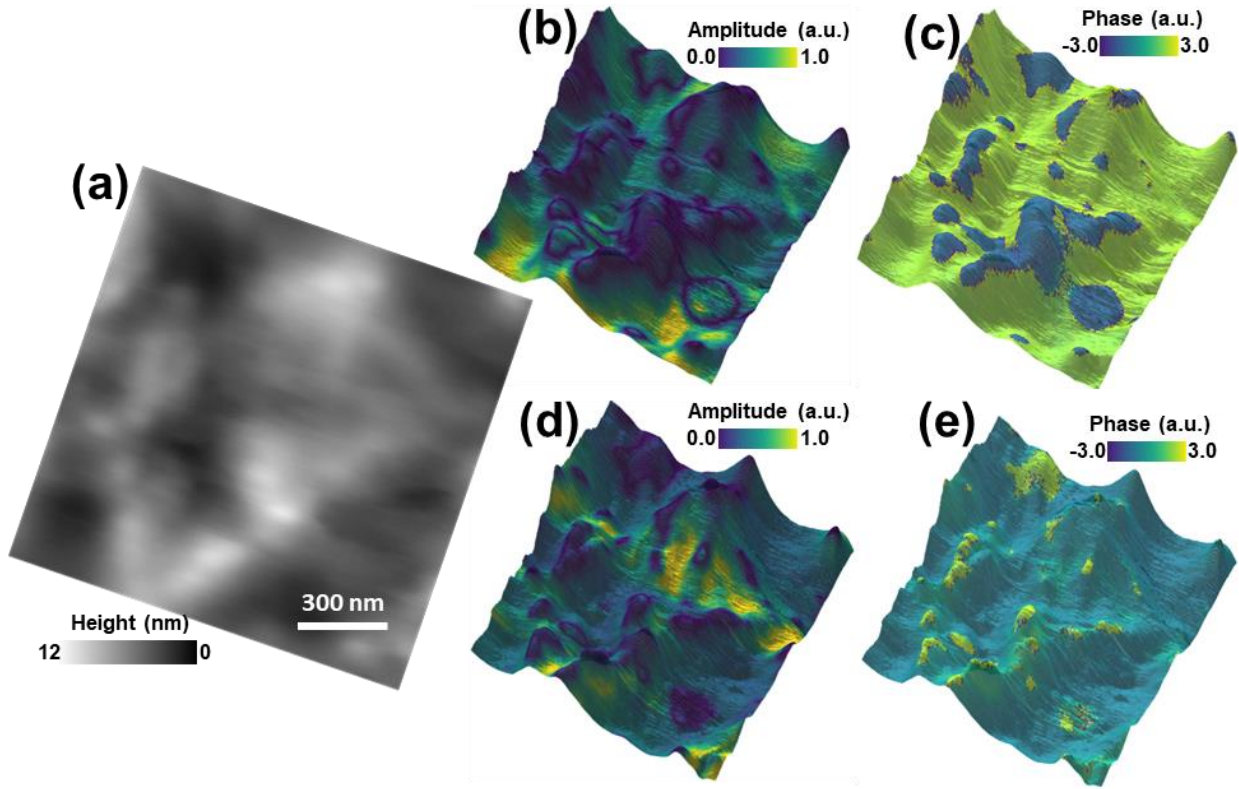

Figure 2. Band Excitation Piezoresponse Force Microscopy shows the relationship between domain evolution and morphology in CIPS upon DC poling. (a) Topography of a CIPS flake. (b-c) BEPFM results of original CIPS flake, (d-e) BEPFM results of the CIPS flake after $V_{dc}$ = 4 V poling. The BEPFM results in (b-e) are plotted over topography to highlight the relationship between domains and morphology, where the color represents PFM amplitude or phase and the geometry represents topography. It is shown in (d-e) that the domain walls move towards the high curvature regions after poling.

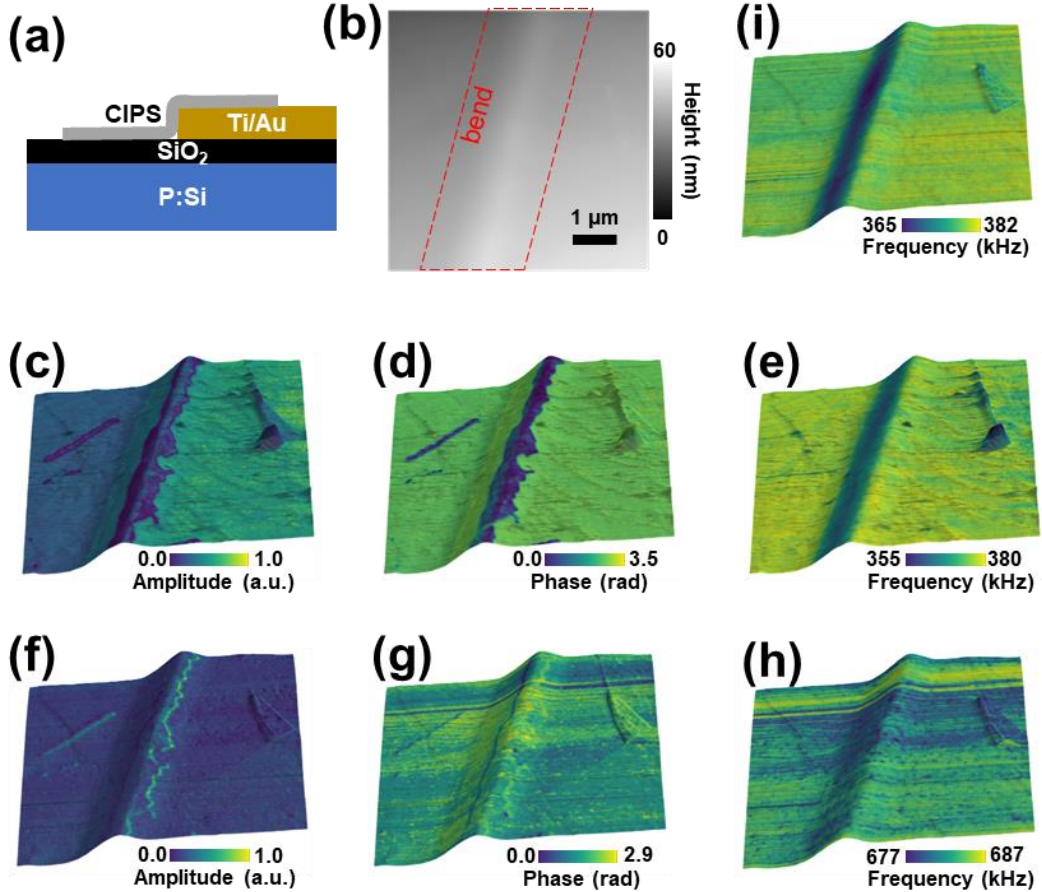

Figure 3. Artificial domains induced by bending. (a) device structure for bending CIPS flake. (b) AFM topography shows the bend structure. (c-e) vertical BEPFM amplitude, phase, and frequency showing the domain structure induced by bending. (f-h) lateral BEPFM amplitude, phase, and frequency showing the in-plane response of bending-induced domains. (i) bluelaser-driven CR-AFM showing the mechanical property of bending-induced domains, the resonant frequency variation at the bend regions is lower by 20 kHz, indicating the material here is softer.

The domain wall motion toward high curvature area induced by mechanical and electric modulation motivated us to explore the correlation between curvature and ferroelectric domains in CIPS flakes. We intentionally introduced curvature in CIPS flakes by transferring the flakes on the edge of the bottom electrodes, thus, the CIPS flake at the edge forms a large curvature, as shown in Figure 3a. The AFM topography result in Figure 3b confirms the curvature formed near the edge area. Next, BEPFM image and spectroscopy measurements were performed to investigate the ferroelectricity near the large curvature area. Shown in Figure 3c-e and Figure 3f-h are the out-of-plane and in-plane BEPFM results, respectively. The results in Figure 3c-d shows a domain along the down-bending edge, while no domain is seen in the up-bending edge. We ascribe this behavior as that the down-bending reverses the polarization. In-plane BEPFM amplitude in Figure 3f shows a slightly higher amplitude response corresponding to the domain walls, which may suggest an in-plane polarization at domain walls. Then, the corresponding resonance frequency map in Figure 3e indicates a frequency shift over the curvature, this frequency shift is further confirmed by a blue-laser driven contact resonance AFM measurement (in contrast to electric driven in PFM) shown in Figure 3i. The frequency shift is relatively small, usually a resonance frequency shift is associated with the local mechanical stiffness change. However, it is worth mentioning that the frequency shift area does not correspond to the ferroelectric domain here, where the domains formed at the down-bending edge but the frequency shift happens at the titled area (probably owing to tip-sample contact variation).

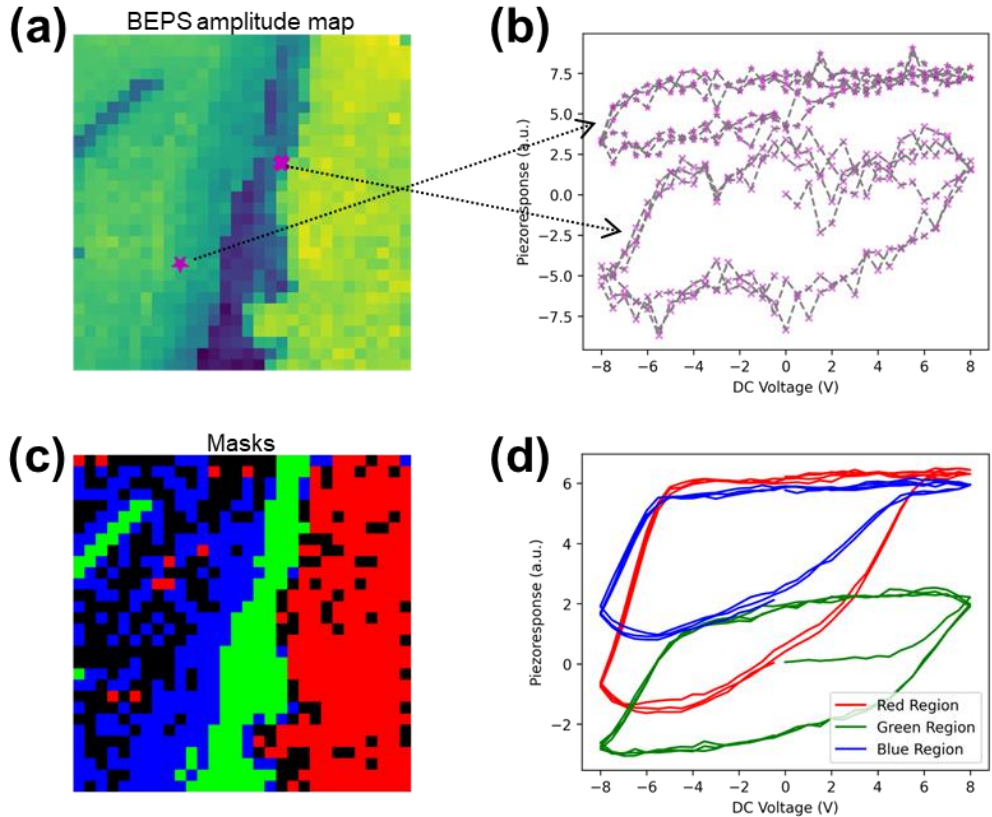

Figure 4. Band Excitation Piezoresponse Spectroscopy (BEPS) results of the artificial bending domains. (a) Amplitude image shows the domain structure. (b) Two example BEPS spectroscopy from the marked locations in (a). (c) Masks correspond to the different domain regions. (d), Averaged BEPS from the corresponding regions in (c).

Band excitation piezoresponse spectroscopy (BEPS) measurement was further performed to explore the polarization dynamics in this area. In BEPS, a DC pulse ($V_{dc}$) is applied to switch the local polarization. Here we note that the electrically driven polarization switching in CIPS is challenging,[34] because the application of electric field often induces ion migration and sample damages. Nevertheless, we investigate the deformation induced by Vdc + Vac). We applied a triangular DC waveform with a magnitude of 8 V to the sample. Shown in Figure 4a is the bending-induced domain structures of the BEPS measurement area and shown in Figure 4b are two exampliers BEPS piezoresponse vs. Vdc loops, indicating the spatial variation of BEPS response in the measurement area. To better understand the BEPS response of the bending-induced domain, we created three masks (as shown in Figure 4c) to mask out the BEPS results corresponding to the high curvature area (green mask) and no-curvature areas (blue and red masks). The averaged BEPS loops corresponding to the masked areas are shown in Figure 4d. it is obvious that the BEPS loops of the high curvature area shows a distinct shape compared to no-curvature area.

To get insight into the mechanisms of the observed phenomena, we performed finite element modelling (FEM) to get insight into the mechanisms of the observed phenomena. Since the bending changes the shape of CIPS layer, FEM results are presented in the local coordinate frame linked with the instant position of crystallographic axes inside the bended layer. The local normal axis $x_n$ coincides with $z$, and the tangential axis $x_t$ coincides with $x$ for the flat layer (see Figure 5).

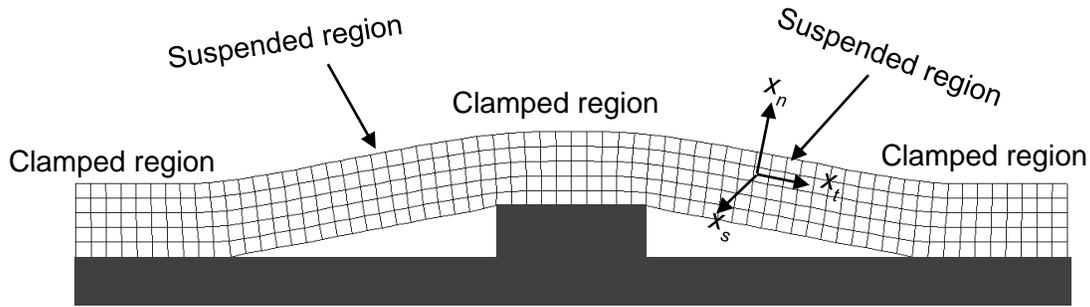

Figure 5. Typical mesh in the bent CIPS nanoflake on the patterned substrate, which is suspended between the patterns and clamped in the regions of the pattern "tops". The pattern lateral size is 50 nm, height varies from 0 to 5 nm, and their period is 350 nm.

Distribution of the normal and tangential polarization components, $P_n$ and $P_t$, inside the flat 5-nm thick CIPS layer suspended between the patterns is shown in the top rows of Figure A2(a) and A2(b), respectively. Since the top and bottom surfaces of the layer are regarded perfectly screened, the single-domain ferrielectric state with the polarization $P_n \sim 3$ µC/cm² (abbreviated as "FI1") is stable in the flat or slightly bended layer. An isostructural transition from the FI1 state to the ferrielectric state with lower polarization $P_n \sim 0.5$ µC/cm² (abbreviated as "FI2") occurs with the bending increase. The FI1-FI2 boundary is relatively sharp (~2.5 nm) because the interface between the FI1 and FI2 regions is almost unchanged. Due to the absence of the uncompensated bound charge, the depolarization electric field is absent inside the flat layer, and negligibly small at the flat FI1-FI2 boundary (see the top rows in Figure A3(a) and A3(b), respectively).

With an increase in the step height below the critical value, the layer bending increases, and well-localized features of polarization, field and strains appear in the inflection points [see Figure 6(a) and Figure A2-A4]. When the pattern height (and so the bending) exceeds the critical value, the FI2 regions rapidly growth and fill in the inflection regions, where the layer is locally flat, as shown in Figure 6(b), and Figure A2-A4. When the pattern height increases further, the FI2 regions expand, and, simultaneously, the regions of FI1 states begin to shrink. Thus, the critical bending of the layer creates the inflected regions, where the FI2 states are stable, and significantly increases the normal polarization component up to 4 µC/cm² in all layer except for the inflection regions with small $P_n \sim 0.4$ µC/cm². The thin boundaries between the FI1 and FI2 states have a small and sign-alternating $P_t$ (about 0.2 µC/cm²), which reaches a maximum at the boundaries.

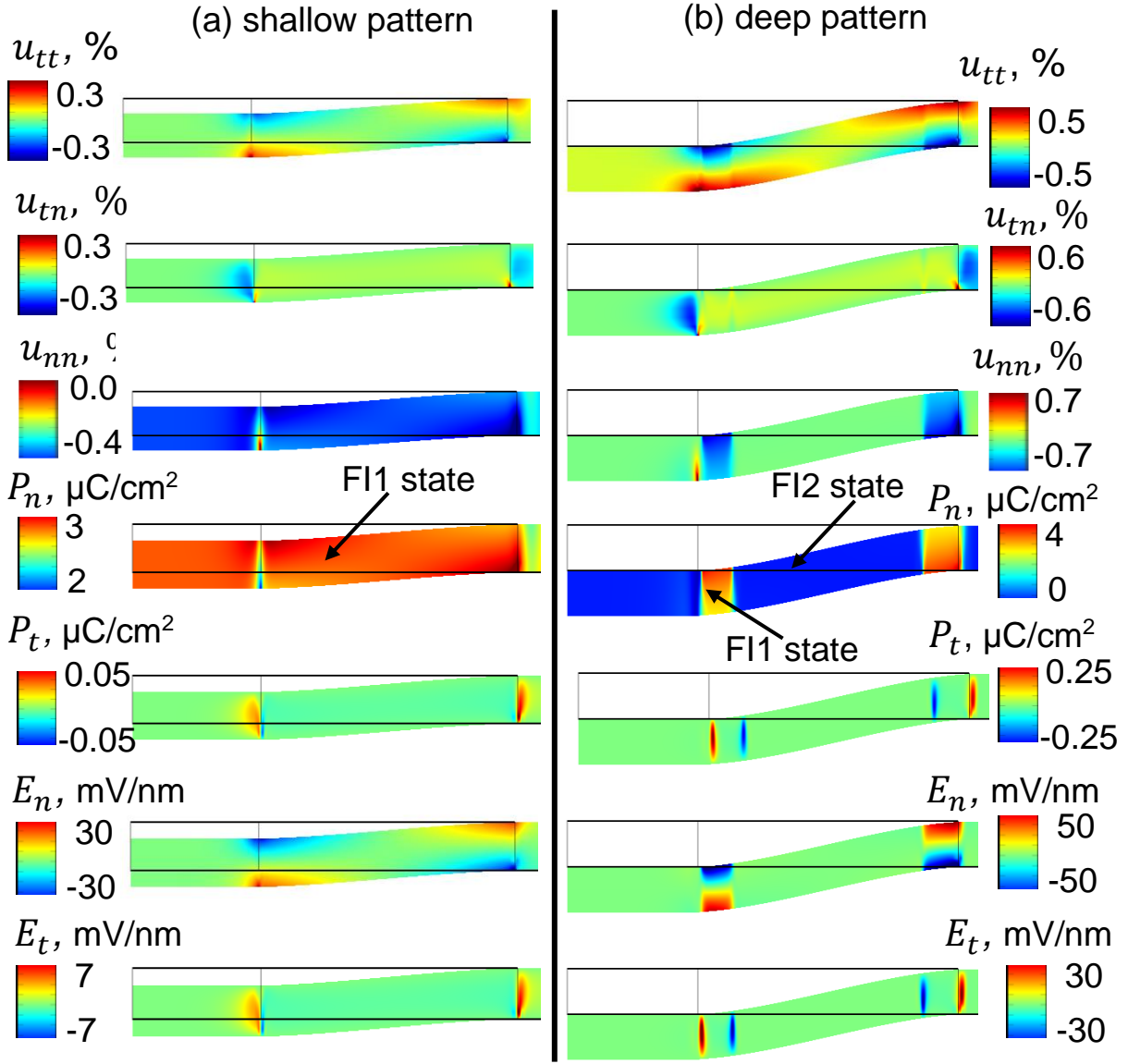

Figure 5. Correlations between the distributions of elastic strains, $u_{tt}$, $u_{tn}$ and $u_{nn}$, polarization components, $P_n$ and $P_t$, and the field components, $E_n$ and $E_n$, in the suspended CIPS layer for different height of patterning, namely 1.7 nm (a) and 5 nm (b). Thin black lines represent initial shape of the flat layer which has an actual shape of rectangle with the 5 nm height of and 155 nm wide. The top and bottom surfaces of the layer are perfectly screened; calculations are performed at 293 K. Color scales show the strain components in %, polarization components in μC/cm$^2$, and field components in mV/nm.

To provide further insight, we develop the mesoscopic Landau-Ginsburg-Devonshire (LGD) theory for slightly bended CIPS layers. We consider different contributions to the static electric polarization of corrugated CIPS film, which is a semiconducting, flexoelectric, piezoelectric and uniaxial ferrielectric low-dimensional material. Let us also assume that the bending is very small, and so $P_n \approx P_3$ in the first approximation.

For a ferroelectric with the one-component out-of-plane spontaneous polarization $P_3$, the bulk density of the LGD functional $F$ depends on $P_3$, elastic stress $\sigma_i$, and their gradients, and has the following form:[27,35]

$$F = \frac{\alpha}{2}P_3^2 + \frac{\beta}{4}P_3^4 + \frac{\gamma}{6}P_3^6 + \frac{\delta}{8}P_3^8 + \frac{g}{2}\left(\frac{\partial P_3}{\partial x_i}\right)^2 - P_i E_i - Q_{i3}\sigma_i P_3^2 - Z_{i33}\sigma_i P_3^4 - \frac{F_{ii3j}}{2}\left(P_3 \frac{\partial \sigma_i}{\partial x_j} - \sigma_i \frac{\partial P_3}{\partial x_j}\right) - \frac{s_{ii}}{2}\sigma_i^2 - s_{ij}\sigma_i\sigma_j.$$

(1)

According to Landau theory,[36] the coefficient $\alpha$ linearly depends on the temperature $T$ for proper ferroics, $\alpha(T) = \alpha_T(T - T_C)$, where $T_C$ is the Curie temperature. All other coefficients in Eq.(1) are supposed to be temperature independent. The coefficient $\delta \geq 0$ for the stability of the free energy for all $P_3$ values. The gradient coefficient $g$ determines the magnitude of the gradient energy. $E_i$ is an electric field. The values $Q_{ijkl}$ and $Z_{ijkl}$ are the linear and nonlinear electrostriction stress tensor components, respectively. $F_{ijkl}$ is the flexoelectric stress tensor[37] determined by the microscopic properties of the material.[38,39] Einstein summation over repeated indexes is used hereinafter. The values $T_C$, $\alpha_T$, $\beta$, $\gamma$ and $\delta$, $Q_{ijkl}$ and $Z_{ijkl}$, given in Table AI in Appendix A, were defined in Ref.[26].

The electric field $E_i$ is expressed via the potential $\varphi$, as $E_i = -\frac{\partial \varphi}{\partial x_i}$. The potential satisfies the Poisson equation, and we assume that $|e\varphi| \ll k_B T$, and introduce the Debye-Hukkel screening length, $L_D = \sqrt{\frac{k_B T \varepsilon_0 \varepsilon_b}{2e^2 n_0}}$, where $\varepsilon_0$ is a universal dielectric constant, $\varepsilon_b$ is the background dielectric constant,[40] $n_0$ is the free carrier density in the unstrained material at $\varphi = 0$ (see Appendix B for details).

Note that the surface corrugation can be expanded in Fourier series, whose each term can induce the dependence of the harmonic displacement of the CIPS layer along X- or Y-axis. For the sake of simplicity, only one-component elastic field is considered, and only one-component of the stress tensor, $\sigma_3$, is assumed to be nonzero and modulated in the direction $x$:

$$\sigma_3(x) = \sigma_3^m \cos(kx). \tag{2}$$

Here $\sigma_3^m$ is the amplitude and $\frac{2\pi}{k}$ is the stress period. Allowing for the Debye-Hukkel approximation, the variation of the free energy (1) yields:

$$(\alpha - 2Q_{ij33}\sigma_{ij})P_3 + (\beta - 4Z_{ij33}\sigma_{ij})P_3^3 + \gamma P_3^5 + \delta P_3^7 - g\frac{\partial^2 P_3}{\partial x_i^2} = E_3 + F_{ii3j}\frac{\partial \sigma_i}{\partial x_j}. \quad (3)$$

Under the condition $k^2 L_D^2 \ll 1$, valid hereinafter due to the high electric conductivity of narrow-gap CIPS, the depolarization contribution can be small enough.

From Eq.(3), that the local change of Curie temperature can be derived as,

$$T_{LC}(\vec{x}) = T_C + \frac{2}{\alpha_T}Q_{33}\sigma_3(x) - \frac{k^2}{\alpha_T\varepsilon_0(\varepsilon_b k^2 + L_D^{-2})}. \quad (4)$$

However, the "local" expression has very little relation to the dependence of the phase transition temperature, $T_{pt}$, as a function of average curvature, which can be measured experimentally. The latter is a "global" coordinate-independent value, because the phase transition may occur in a spatial region much bigger than the period $\frac{2\pi}{k}$ of a strain/corrugation. In this case $T_{pt} \approx T_C$, because the average value $\langle \sigma_3(\vec{x}) \rangle \approx 0$. However, when the lateral size $L \sim \frac{2\pi}{k}$ of a bended region becomes compatible or higher than the characteristic length of polarization correlations, $L_c$, the local phase transition may occur in the region.

Using the decoupling approximation,[41,42] the vertical displacement $U_i$ of CIPS surface, can be estimated under the PFM tip as:[43]

$$U_i(\vec{y}) \approx c_{kjmn}\int_{-\infty}^{\infty}d\xi_1\int_{-\infty}^{\infty}d\xi_2\int_0^h d\xi_3 \frac{\partial G_{im}(-\xi_1,-\xi_2,\xi_3)}{\partial \xi_n}E_l(\vec{\xi})d_{lkj}(\vec{y}-\vec{\xi}), \quad (5)$$

where $G_{im}(-\xi_1,-\xi_2,\xi_3)$ is a Green function, $E_l(\vec{\xi})$ is an electric field induced by the PFM probe in the CIPS layer. The piezoelectric coefficients $d_{lkj}(\vec{x})$ are proportional to the spontaneous polarization and dielectric susceptibility:[44]

$$d_{ijk}(\vec{x}) \approx 2\varepsilon_0\chi\delta_{km}Q_{ijm3}P_3(\vec{x}), \quad (6)$$

where $\chi$ is the linear dielectric susceptibility, and is a Kronecker-delta symbol. For the considered case of $\sigma_3(x)$ given by Eq.(2), the dependence of $\chi$ on $k$ is analyzed in Figure B4 in Appendix B.

Expressions (5)-(6) imply that the surface displacement $U_i$ is linearly proportional to the integral of $P_3(\vec{x})$, which variation is in turn proportional to the stress (or strain) and its gradient induced by the patterned substrate. For a flattened tip apex and small spontaneous polarization $P_S$ (e.g., near the Curie temperature of CIPS, 293 K), a vertical PFM response, defined as $d_{33}^{eff} = \frac{dU_3}{dV}$, is proportional to the product $\chi P_3(\vec{x})$. In a realistic case of a smooth corrugation and small $L_D$, when $kL_D \ll 1$, we can use an estimation:

$$d_{33}^{eff} \sim Q_{33}\varepsilon_0\chi\left(f_{3jkl} - \frac{\Sigma_{kl}^g}{2e}\right)\frac{\partial}{\partial x_j}u_{kl}(\vec{x}) + 2P_s(q_{lm33} + 2z_{lm33}P_s^2)u_{kl}(\vec{x}). \qquad (7)$$

Here $f_{3jkl}$ is the flexoelectric strain tensor, $q_{lm33}$ and $z_{lm33}$ are the linear and nonlinear electrostriction strain tensor components, respectively; $\Sigma_{kl}^g$ is a deformation potential tensor. Expression (7) shows that the bending can create a smooth domain structure in a "weak" ferrielectric CIPS, and the structure determines the amplitude and phase of PFM response.

Due to the strong, negative and temperature-dependent nonlinear electrostriction couplings ($Z_{i33} < 0$), and the "inverted" signs of the linear electrostriction coupling ($Q_{33} < 0$, $Q_{23} > 0$ and $Q_{13} > 0$) for CIPS, the expected pressure effect on the local polarization switching is complex and unusual in comparison with many other ferroelectrics with $Q_{33} > 0$, $Q_{23} < 0$, and $Q_{13} < 0$.

Results of numerical solution of Eq.(3) are shown in Figure 7 (see also Figures B1-B3 in Appendix B). From Figure 7(a), the local stress changes the polarization profile, namely it causes shallow wells, which exist in tensiled regions at temperatures (250 – 280) K, and deeper wells (up to $P = 0$), which exist in tensiled regions at temperatures (290 – 300) K. Relatively small increase (hills) of the polarization profile exist in compressed regions at temperatures (250 – 300) K. Figure 7(b) demonstrates the strong and influence of periodic stress on the hysteresis loops shape, magnitude of the remanent polarization, and coercive fields (compare blue, black and red loops). The rectangular-shaped loop at compressive stress, and double loops at tensile stress can be explained by the anomalous temperature dependence and "inverted" signs of CIPS linear and nonlinear electrostriction coupling coefficients. By varying the sign of applied stress (from expansion to compression) and its magnitude (from zero to several hundreds of MPa), a quasi-static double hysteresis loops can transform into pinched or single hysteresis loops (compare red, black and blue loops).

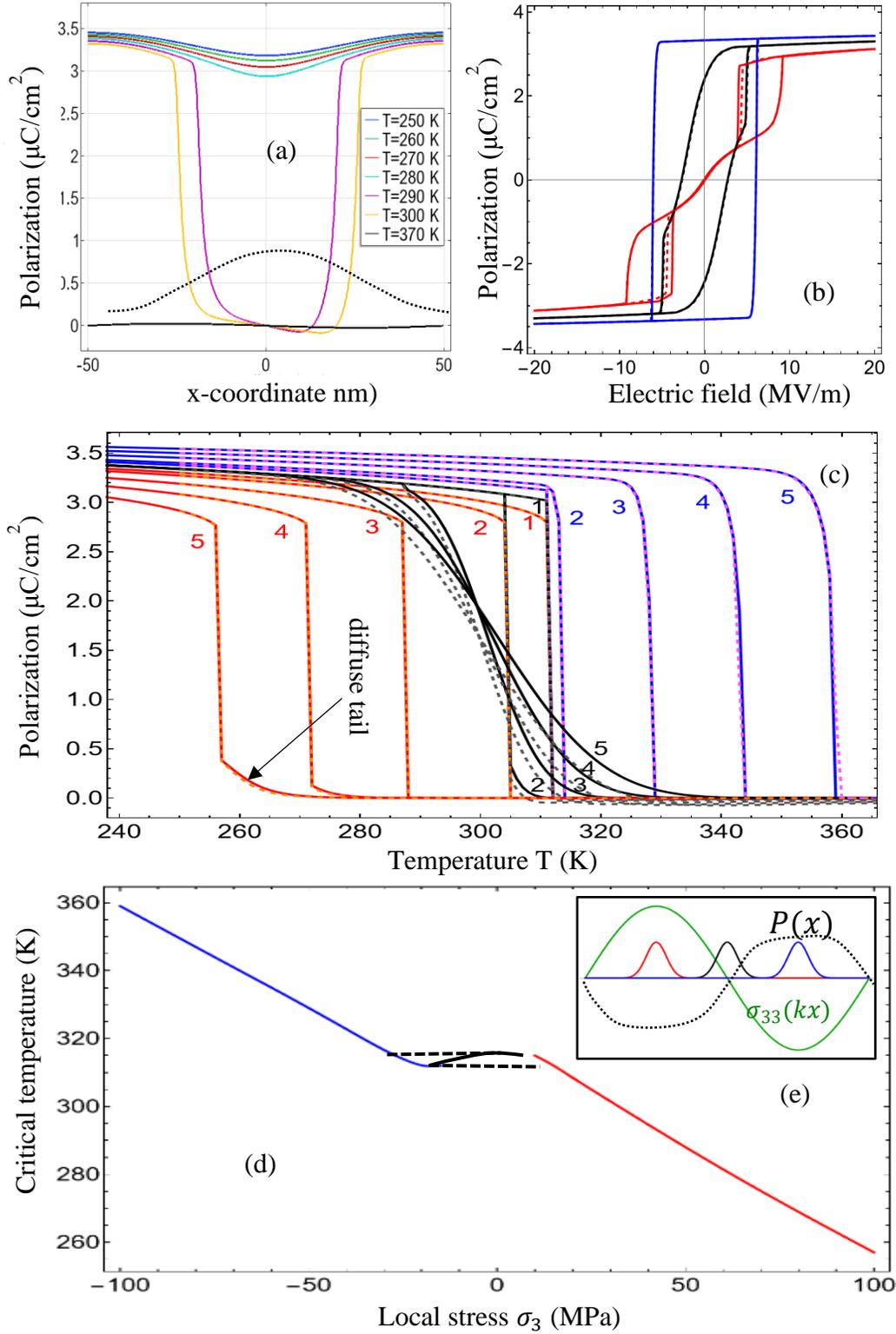

Figure 7. (a) Distribution of spontaneous polarization induced by a co-sinusoidally modulated stress with the amplitude $\sigma_3^m$=50 mpa and period 100 nm calculated for several temperatures from 250 K to 300 K (see legend for description of solid curves colors). Flexoelectric coefficient $f =$

$2 \cdot 10^{-11}$M³/C. Dotted curve is a schematical distribution of the stress distribution. (b) Quasi-static hysteresis loops of local polarization calculated at 298 K. Different loops correspond to the three 10-nm wide regions with maximal stress $\sigma_3(x) = \sigma_3^m$ (red curves), zero stress $\sigma_3(x) = 0$ (black curves), and minimal stress $\sigma_3(x) = -\sigma_3^m$ (blue curves). (c) Temperature dependence of the spontaneous polarization values, which are averaged over three 10-nm wide regions, located near the points with maximal stress $\sigma_3(x) = \sigma_3^m$ (red curves 1-5), zero stress $\sigma_3(x) = 0$ (black curves 1-5), and minimal stress $\sigma_3(x) = -\sigma_3^m$ (blue curves 1-5) with the amplitude $\sigma_3^m$=15, 25, 50, 75 and 100 mpa (curves 1, 2, 3, 4 and 5, respectively). Flexoelectric coefficient $f = 0$ for solid curves and $f = 2 \cdot 10^{-11}$M³/C for dashed curves in the plots (b)-(c). (d) Stress dependence of the critical temperature, defined as the temperature values, where the polarization sharply decreases (diffuse tails are neglected).

From Figures 7(c)-(d), the interaction of a ferrielectric soft mode with a periodic elastic deformation, induced by the patterned substrate, leads to the local shift of the phase transition temperature, and the temperature is higher for compressed regions [see the blue curves in Figure 7(c) and the blue line in Figure 7(d)], and lower for tensiled regions [see the blue curves in Figure 7(c) and the blue line in Figure 7(d)]. The local changes are visible only if the stress period is much higher than the correlation length. The change of the phase transition temperature, averaged over the corrugation period, is equal to zero. This is due to the fact, that the average stress is zero. So, curved, but stress-free regions, look indifferent to the phase transition shift. Polarization averaging over the 10-nm wide region imitates the Gaussian field caused by a 10-nm size of the PFM tip-surface contact, and is schematically shown in Figure 7(e). Polarization ripples, schematically shown by the black dotted curve in the inset, are caused by the modulated stress, shown by the green curve.

The flexoelectric effect induces a relatively small built-in electric field in the right-hand side of Eq.(3), which induces a slight asymmetry of polarization profiles, as shown in Figure 7(a), leads to a slight horizonal shift of local polarization hysteresis, as shown by dashed curves in Figure 7(b), and smears the phase transition point, as shown by dashed curves in Figure 7(c)-(d). The asymmetry and smearing become stronger with increase in the stress amplitude [compare, e.g., Figures B1(c) and B1(d) in Appendix B]. As expected, the role of the flexoelectric effect becomes much stronger near the phase transition point.

To summarize, we explore the effect of bending on ferroelectric domains in CIPS flakes. In PFM experiment, we observed bending induced domains in CIPS film and different polarization-voltage hysteresis loops in the bending regions and non-bending regions. Our simulation results show that bending can create a smooth domain structure in a "weak" ferrielectric

CIPS, and the structure determines the amplitude and phase of PFM response. The local stress changes the polarization profile by causing shallow and wells, which exist in tensile regions at temperatures 250 – 300 K. Relatively small increase (hills) of the polarization profile exist in compressed regions at the same temperatures. The periodic stress shows a strong influence on the hysteresis loops shape (consistent with experimental observation), magnitude of the remanent polarization, and coercive fields. The squire-shaped loop at compressive stress, and double loops at tensile stress can be explained by the anomalous temperature dependence and "inverted" signs of CIPS linear and nonlinear electrostriction coupling coefficients. The local changes are visible only if the stress period is much higher than the correlation length. The change of the phase transition temperature, averaged over the corrugation period, is equal to zero. This is due to the fact, that the average stress is zero. So, curved, but stress-free regions, look indifferent to the phase transition shift. The flexoelectric effect induces a relatively small built-in electric field, which induces a slight asymmetry of polarization profiles, leads to a slight horizonal shift of local polarization hysteresis and smears the phase transition point. The role of the flexoelectric effect becomes much stronger near the phase transition point.

Acknowledgements

This effort (materials synthesis, PFM measurements) was supported as part of the center for 3D Ferroelectric Microelectronics (3DFeM), an Energy Frontier Research Center funded by the U.S. Department of Energy (DOE), Office of Science, Basic Energy Sciences under Award Number DE-SC0021118. The research (PFM measurements) was performed and partially supported at Oak Ridge National Laboratory's Center for Nanophase Materials Sciences (CNMS), a U.S. Department of Energy, Office of Science User Facility, under user proposal CNMS2023-A-01858. The growth of single crystals of CIPS used in this work was supported by the National Science Foundation through the Penn State 2D Crystal Consortium-Materials Innovation Platform (2DCC-MIP) under NSF cooperative agreement DMR-1539916, and DMR-2039351. A.N.M. acknowledges support from the National Research Fund of Ukraine (project "Low-dimensional graphene-like transition metal dichalcogenides with controllable polar and electronic properties for advanced nanoelectronics and biomedical applications", grant application 2020.02/0027). E.A.E. acknowledges support from the National Academy of Sciences of Ukraine.

# References

1. Novoselov, K. S. *et al.* Electric field effect in atomically thin carbon films. *science* **306**, 666-669 (2004).
2. Manzeli, S., Ovchinnikov, D., Pasquier, D., Yazyev, O. V. & Kis, A. 2D transition metal dichalcogenides. *Nature Reviews Materials* **2**, 1-15 (2017).
3. Mayorov, A. S. *et al.* Micrometer-scale ballistic transport in encapsulated graphene at room temperature. *Nano letters* **11**, 2396-2399 (2011).
4. Balandin, A. A. Thermal properties of graphene and nanostructured carbon materials. *Nature materials* **10**, 569-581 (2011).
5. Yankowitz, M. *et al.* Tuning superconductivity in twisted bilayer graphene. *Science* **363**, 1059-1064 (2019).
6. Wang, L. *et al.* Correlated electronic phases in twisted bilayer transition metal dichalcogenides. *Nature materials* **19**, 861-866 (2020).
7. Kerelsky, A. *et al.* Maximized electron interactions at the magic angle in twisted bilayer graphene. *Nature* **572**, 95-100 (2019).
8. Choi, Y. *et al.* Electronic correlations in twisted bilayer graphene near the magic angle. *Nature physics* **15**, 1174-1180 (2019).
9. Huang, B. *et al.* Layer-dependent ferromagnetism in a van der Waals crystal down to the monolayer limit. *Nature* **546**, 270-273 (2017).
10. Belianinov, A. *et al.* CuInP2S6 room temperature layered ferroelectric. *Nano letters* **15**, 3808-3814 (2015).
11. Kalinin, S. V. & Meunien, V. Quantum Flexoelectricity in Low Dimensional Systems. *arXiv preprint arXiv:0707.3971* (2007).
12. Dai, Z., Liu, L. & Zhang, Z. Strain engineering of 2D materials: issues and opportunities at the interface. *Advanced Materials* **31**, 1805417 (2019).
13. Zhang, Y. *et al.* Strain modulation of graphene by nanoscale substrate curvatures: a molecular view. *Nano letters* **18**, 2098-2104 (2018).
14. Yang, W., Chen, S., Ding, X., Sun, J. & Deng, J. Reducing Threshold of Ferroelectric Domain Switching in Ultrathin Two-Dimensional CuInP2S6 Ferroelectrics via Electrical–Mechanical Coupling. *The Journal of Physical Chemistry Letters* **14**, 379-386 (2023).
15. Sun, Z.-Z., Xun, W., Jiang, L., Zhong, J.-L. & Wu, Y.-Z. Strain engineering to facilitate the occurrence of 2D ferroelectricity in CuInP2S6 monolayer. *Journal of Physics D: Applied Physics* **52**, 465302 (2019).
16. Chen, C. *et al.* Large-scale domain engineering in two-dimensional ferroelectric CuInP2S6 via giant flexoelectric effect. *Nano Letters* **22**, 3275-3282 (2022).
17. Rahman, S., Yildirim, T., Tebyetekerwa, M., Khan, A. R. & Lu, Y. Extraordinary Nonlinear Optical Interaction from Strained Nanostructures in van der Waals CuInP2S6. *ACS nano* **16**, 13959-13968 (2022).
18. Ming, W. *et al.* Flexoelectric engineering of van der Waals ferroelectric CuInP2S6. *Science Advances* **8**, eabq1232 (2022).
19. Liu, F. *et al.* Room-temperature ferroelectricity in CuInP2S6 ultrathin flakes. *Nature communications* **7**, 1-6 (2016).
20. Brehm, J. A. *et al.* Tunable quadruple-well ferroelectric van der Waals crystals. *Nature Materials* **19**, 43-48 (2020).
21. Morozovska, A. N., Eliseev, E. A., Kelley, K. & Kalinin, S. V. Temperature-Assisted Piezoresponse Force Microscopy: Probing Local Temperature-Induced Phase Transitions in Ferroics. *Physical Review Applied* **18**, 024045 (2022).

22  Osada, M. & Sasaki, T. The rise of 2D dielectrics/ferroelectrics. *APL Materials* **7**, 120902 (2019).
23  Wu, M. & Jena, P. The rise of two-dimensional van der Waals ferroelectrics. *Wiley Interdisciplinary Reviews: Computational Molecular Science* **8**, e1365 (2018).
24  Susner, M. A. *et al.* Cation–eutectic transition via sublattice melting in CuInP2S6/In4/3P2S6 van der Waals layered crystals. *ACS nano* **11**, 7060-7073 (2017).
25  Tolédano, P. & Guennou, M. Theory of antiferroelectric phase transitions. *Physical Review B* **94**, 014107 (2016).
26  Morozovska, A. N., Eliseev, E. A., Kalinin, S. V., Vysochanskii, Y. M. & Maksymovych, P. Stress-induced phase transitions in nanoscale Cu In P 2 S 6. *Physical Review B* **104**, 054102 (2021).
27  Bourdon, X., Maisonneuve, V., Cajipe, V., Payen, C. & Fischer, J. Copper sublattice ordering in layered CuMP2Se6 (M= In, Cr). *Journal of alloys and compounds* **283**, 122-127 (1999).
28  Susner, M. A., Chyasnavichyus, M., McGuire, M. A., Ganesh, P. & Maksymovych, P. Metal thio- and selenophosphates as multifunctional van der Waals layered materials. *Advanced Materials* **29**, 1602852 (2017).
29  Maisonneuve, V., Cajipe, V., Simon, A., Von Der Muhll, R. & Ravez, J. Ferrielectric ordering in lamellar CuInP 2 S 6. *Physical Review B* **56**, 10860 (1997).
30  Banys, J. *et al. Van Der Waals Ferroelectrics: Properties and Device Applications of Phosphorous Chalcogenides*. (John Wiley & Sons, 2022).
31  Niu, L. *et al.* Controlled synthesis and room-temperature pyroelectricity of CuInP2S6 ultrathin flakes. *Nano Energy* **58**, 596-603 (2019).
32  Xu, D.-D. *et al.* Ion adsorption-induced reversible polarization switching of a van der Waals layered ferroelectric. *Nature communications* **12**, 655 (2021).
33  Balke, N. *et al.* Locally controlled Cu-ion transport in layered ferroelectric CuInP2S6. *ACS applied materials & interfaces* **10**, 27188-27194 (2018).
34  Zhou, S. *et al.* Anomalous polarization switching and permanent retention in a ferroelectric ionic conductor. *Materials Horizons* **7**, 263-274 (2020).
35  Morozovska, A. N., Eliseev, E. A., Vysochanskii, Y. M., Khist, V. V. & Evans, D. R. Screening-induced phase transitions in core-shell ferroic nanoparticles. *Physical Review Materials* **6**, 124411 (2022).
36  Landau, L. D., Lifšic, E. M., Lifshitz, E. M., Kosevich, A. M. & Pitaevskii, L. P. *Theory of elasticity: volume 7*. Vol. 7 (Elsevier, 1986).
37  Yudin, P. & Tagantsev, A. Fundamentals of flexoelectricity in solids. *Nanotechnology* **24**, 432001 (2013).
38  Mashkevich, V. & Tolpygo, K. Electrical, optic and elastic properties of crystals of diamond type. *Sov. Phys.-JETP* **4**, 455-460 (1957).
39  Kogan, S. M. Piezoelectric effect during inhomogeneous deformation and acoustic scattering of carriers in crystals. *Soviet Physics-Solid State* **5**, 2069-2070 (1964).
40  Tagantsev, A. K., Cross, L. E. & Fousek, J. *Domains in ferroic crystals and thin films*. Vol. 13 (Springer, 2010).
41  Felten, F., Schneider, G., Saldaña, J. M. & Kalinin, S. Modeling and measurement of surface displacements in BaTiO 3 bulk material in piezoresponse force microscopy. *Journal of Applied Physics* **96**, 563-568 (2004).
42  Scrymgeour, D. A. & Gopalan, V. Nanoscale piezoelectric response across a single antiparallel ferroelectric domain wall. *Physical Review B* **72**, 024103 (2005).
43  Morozovska, A. N., Eliseev, E. A., Bravina, S. L. & Kalinin, S. V. Resolution-function theory in piezoresponse force microscopy: Wall imaging, spectroscopy, and lateral resolution. *Physical Review B* **75**, 174109 (2007).
44  Smolensky, G.    (New York-London-Paris-Montreux-Tokyo, 1984).